\documentclass[preprint,12pt]{elsarticle}

\usepackage{amssymb,amsmath,amsthm,bm}
\usepackage{ dsfont }
\usepackage{color}

\newcommand{\Prob} {\mbox{$\rm{Prob}$\,}}

\usepackage{lineno}

\journal{Theoretical Population Biology}

\begin{document}

\begin{frontmatter}



\title{Stationary distribution of a 2-island 2-allele Wright-Fisher diffusion model with slow mutation and migration rates}

\author[label1,label2]{Conrad J.\ Burden}
\ead{conrad.burden@anu.edu.au}
\author[label3]{Robert C.\ Griffiths}
\ead{griff@stats.ox.ac.uk}
\address[label1]{Mathematical Sciences Institute, Australian National University, Canberra, Australia}
\address[label2]{Research School of Biology, Australian National University, Canberra, Australia}
\address[label3]{Department of Statistics, University of Oxford, UK}

\begin{abstract}
The stationary distribution of the diffusion limit of the 2-island, 2-allele Wright-Fisher with small but otherwise 
arbitrary mutation and migration rates is investigated.  Following a method developed by~\cite{burden2016approximate,burden2017rate}
for approximating the forward Kolmogorov equation, 
the stationary distribution is obtained to leading order as a set of line densities on the edges of the sample space, corresponding to 
states for which one island is bi-allelic and the other island is non-segregating, and a set of point masses at the corners of the sample space, 
corresponding to states for which both islands are simultaneously non-segregating.  Analytic results for the corner probabilities and line densities are verified 
independently using the backward generator and for the corner probabilities using the coalescent.  
\end{abstract}


\end{frontmatter}



\section{Introduction}
\label{sec:Introduction}

Island models of migration between partially isolated subpopulations date to the pioneering work of~\cite{wright1943isolation}.  Generalisations of 
Wright's original model have subsequently found applications in genetics, ecology and linguistics~\citep{blythe2007stochastic}.  The current paper 
deals with the diffusion limit of a Wright-Fisher model of a finite number of subpopulations undergoing migration and neutral mutations between a finite number of 
allele types~\citep{de2004importance}.  

The stationary distribution of this model for arbitrary migration and mutation rate matrices remains unknown, even for the simplest non-trivial case of two islands 
and two allele types.  Here we investigate the stationary distribution in the limit of small scaled migration and mutation rates using a method developed 
by~\cite{burden2016approximate,burden2017rate}, which was in turn inspired by the boundary mutation models developed by \cite{vogl2015inference} 
and \cite{schrempf2017alternative} in the context of Moran models.  The method relies on the fact that for low migration and mutation rates 
the dynamics is strongly dominated by genetic drift over most of the sample space $\Omega$ of the distribution, except near the boundary of $\Omega$.  
Thus the role of mutation and migration is, 
at lowest order, to provide boundary conditions for a forward Kolmogorov equation in which only drift is relevant.  To place the argument on a more 
rigorous footing, we use a series expansion to obtain the leading order behaviour of the stationary distribution near the boundary of $\Omega$.  

For the case of muti-allelic neutral diffusion in a single population, $\Omega$ is a simplex.  Burden and Tang show that in this case the stationary distribution can be conveniently 
expressed as a set of point masses at the corners of $\Omega$ corresponding to the relative probability of alleles at non-segregating sites and line densities 
on the edges of $\Omega$ corresponding to the site frequencies of bi-allelic polymorphisms.  This purpose of this paper is primarily to demonstrate that the method 
can be carried over to subdivided population models by adapting the method to the 2-island, 2-allele Wright-Fisher model, for which the sample space is 
$\Omega = [0, 1] \times [0, 1]$.  

More recently, \cite{BurdenGriffiths18} have recreated results of \cite{burden2016approximate} for multi-allelic neutral Wright-Fisher diffusion using 
approaches based on the generator of the backward Kolmogorov equation and on the coalescent.  A second aim of the current paper is to confirm our results 
for the 2-island, 2-allele Wright-Fisher model using analogous backward-generator and coalescent methods.  

The structure of the paper is as follows: The discrete Wright-Fisher island model is defined and the corresponding forward-Kolmogorov equation for the 2-island, 2-allele 
case is given in Section~\ref{sec:WHIslandModel}.  
Section~\ref{sec:AlternativeAppr} is a brief review of existing related work before embarking on our approach.
Series expansions near the boundary are used to find the approximate stationary distribution terms of effective corner 
probabilities and effective line densities in Section~\ref{sec:StationaryDist}.  In Sections~\ref{sec:AlternativeFromBackward} and \ref{sec:AlternativeFromCoalescent} 
results are confirmed from independent derivations based on the backward generator and the coalescent respectively.  
Section~\ref{sec:ComparisonNumerical} compares the theory with a numerical simulation of the discrete model.  Conclusions are 
drawn in Section~\ref{sec:Conclusions}.

%
%
\section{The Wright-Fisher island model}
\label{sec:WHIslandModel}

The Wright-Fisher model with $K$ alleles and $g$ islands with haploid populations $M_1, \ldots , M_g$ is defined by the 
Markov transition matrix 
\begin{eqnarray}
\lefteqn{\Prob(\mathbf{Y}(\tau + 1) = \mathbf{y}(\tau + 1), \mathbf{Y}(\tau) = \mathbf{y}(\tau))  }  \nonumber \\
	 &  = & \begin{cases} 
		\displaystyle \prod_{i = 1}^g \left[ \frac{M_i !}{\prod_{a = 1}^K y_{ia}!} \prod_{a = 1}^K \psi_{ia} (\mathbf{y}(\tau))^{y_{ia}(\tau + 1)} \right] \\ \\
			\qquad\qquad\qquad \text{if } \sum_{a = 1}^K y_{ja} = M_j \text{ for each } j = 1, \ldots, g, \\ \\
		0 	\qquad\qquad\qquad \text{otherwise}. 
	\end{cases}			\label{transitionMatrix}
\end{eqnarray}
Here $Y_{ia}(\tau)$ is the number of individuals of allele type $A_a$ on island $i$ at discrete time step $\tau = 0, 1, 2, \ldots$, and  $\psi_{ia} (\mathbf{y})$ 
is the probability that any given individual on island $i$ in generation $\tau + 1$ is born as allele type $A_a$ given the configuration  $\mathbf{y}$ at time 
step $\tau$.  More specifically, consider a neutral model with with mutation rates from allele $A_a$ to allele $A_b$ per generation of $u_{ab}$ and migration 
rates defined by a probability $v_{ij}$ that an individual on island $i$ is the offspring of an individual from island $j$ from the previous generation, where 
\begin{equation}
u_{ab}, v_{ij} \ge 0, \qquad \sum_{b = 1}^K u_{ab} = \sum_{j = 1}^g v_{ij} = 1,  
\end{equation}
for $a = 1, \ldots, K$; $i = 1, \ldots, g$.   Then 
\begin{equation}
\psi_{ia} (\mathbf{y}) = \sum_{j = 1}^g v_{ij} \sum_{b = 1}^K \frac{y_{jb}}{M_j} u_{ba}. 
\end{equation}

From here on we consider the diffusion limit of the case of $g = 2$ islands and $K = 2$ alleles.  A diffusion limit consistent with the usual coalescent 
time~\citep{herbots1997structured}  is obtained by defining a continuum time $t$ 
and relative type-$A_1$ allele frequencies $X_1(t)$ and $X_2(t)$ on islands 1 and 2 respectively by 
\begin{equation}
t = \frac{\tau}{M}, \quad X_{i}(t) = \frac{1}{M_i} Y_{i1}(\tau) = 1 - \frac{1}{M_i} Y_{i2}(\tau),		\label{tAndXDefn}
\end{equation}
where 
\begin{equation}
M = M_1 + M_2.
\end{equation}  
We also define mutation and migration rates per unit continuum time by 
\begin{equation}
\begin{split}
q_{12} = Mu_{12} = M(1 - u_{11}), \qquad q_{21} = Mu_{21} = M(1 - u_{22}), \\
m_{12} = Mv_{12} = M(1 - v_{11}), \qquad m_{21} = Mv_{21} = M(1 - v_{22}), 
\end{split}		\label{mAndQDefn}
\end{equation}		
and relative island population sizes by 
\begin{equation}
r_1 = \frac{M_1}{M}, \qquad r_2 = \frac{M_2}{M}.  \label{relPopSizes}
\end{equation}

\begin{figure}[t!]
\begin{center}
\centerline{\includegraphics[width=\textwidth]{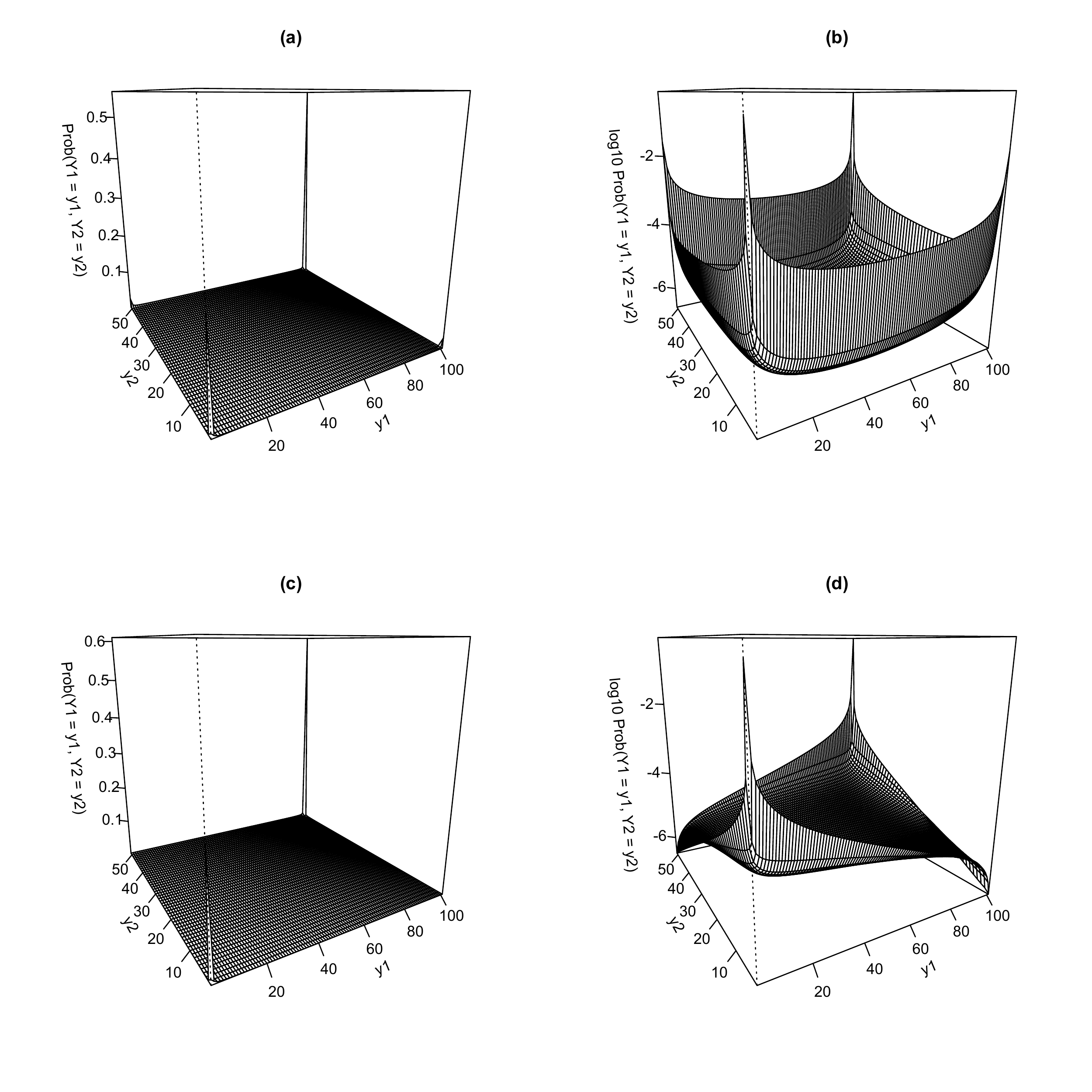}}
\caption{Stationary distribution of a 2-island, 2-allele model with low mutation rates.  
The abundances of type-$A_1$ alleles on island-1 and island-2 are $y_1$ and $y_2$ respectively.
Parameters for plots (a) and (b) are $M_1 = 100$, $M_2 = 50$, $u_{12} = 0.5 \times 10^{-4}$, $u_{21} = 1 \times 
10^{-4}$, $v_{12} = 8 \times 10^{-4}$, $v_{21} = 0.2 \times 10^{-4}$.
Parameters for plots (c) and (d) are the same except $v_{12} = v_{21} = 1 \times 10^{-2}$.
The vertical scales of plots (a) and (c) are linear and of plots (b) and (d) logarithmic to base $10$.  } 
\label{fig:StationaryPersp}
\end{center}
\end{figure}

The corresponding forward-Kolmogorov equation for the density $f(x_1, x_2; t)$ of the joint distribution of $X_1(t)$ and $X_2(t)$ 
is~\citep[see for example][Eq.~(90)]{blythe2007stochastic}
\begin{eqnarray}
\frac{\partial f(x_1, x_2; t)}{\partial t} & = & \frac{1}{2} \sum_{i = 1}^2 \frac{1}{r_i} \frac{\partial^2}{\partial x_i^2} \left\{ x_i(1 - x_i) f(x_1, x_2; t) \right\} \nonumber \\
	&  & - \frac{\partial}{\partial x_1} \left[ \left\{ m_{12}(x_2 - x_1) - x_1 q_{12} + (1 - x_1) q_{21} \right\} f(x_1, x_2; t) \right] \nonumber \\
	&  & - \frac{\partial}{\partial x_2} \left[ \left\{ m_{21}(x_1 - x_2) - x_2 q_{12} + (1 - x_2) q_{21} \right\} f(x_1, x_2; t) \right], \nonumber \\ \label{forwardKolmog2}
\end{eqnarray}
for $(x_1, x_2) \in \Omega = [0, 1] \times [0, 1]$, and $t \ge 0$.  No analytic solution is known for this equation for arbitrary parameters $q_{ab}$, $m_{ij}$ and $r_i$, 
even for the stationary distribution.  

Numerical solutions to the stationary distribution of the discrete model with transition matrix Eq.~(\ref{transitionMatrix}) for a case with $g = 2$ islands, $K = 2$ alleles 
and low scaled mutation rates $q_{12}$ and $q_{21}$ are 
shown in Fig.~\ref{fig:StationaryPersp}.  In subfigures (a) and (b) the scaled mutation and migration rates 
\begin{equation}
\begin{split}
q_{12} = 0.75 \times 10^{-2}, \qquad q_{21} = 1.5 \times 10^{-2}\\
m_{12} = 12 \times 10^{-2}, \qquad m_{21} = 0.3 \times 10^{-2},  
\end{split}		\label{numericScaledParametersA}
\end{equation}
are all $<< 1$.  In this case the stationary distribution is strongly concentrated at the corners and the boundary.  More specifically, 
for this particular simulation we observe that alleles are likely to be fixed on both islands simultaneously, the most likely configurations being those for 
which both islands are fixed for the same allele.  Configurations for which both islands are simultaneously segregating are extremely unlikely.  
It is this situation which we consider in this paper.

In subfigures (c) and (d) the scaled mutation and migration rates  
\begin{equation}
q_{12} = 0.75 \times 10^{-2}, \qquad q_{21} = 1.5 \times 10^{-2}, \qquad m_{12} = m_{21} = 1.5,  
\label{numericScaledParameters2}
\end{equation}
are such that the scaled mutation rates are small, but the scaled migration rates are not.  In this case the two islands' populations are 
more closely coupled, and the probability that  a site can be simultaneously segregating on both islands is first order in the mutation rates.  In other words, 
the stationary distribution is not strongly confined to the boundary of $\Omega$.  This situation is not considered in this paper.  
%
%
\section{Alternative approaches}
\label{sec:AlternativeAppr}

The approach taken in this paper is to consider the diffusion limit, defined by Eq.~(\ref{tAndXDefn}), in a situation where the 
off-diagonal scaled mutation and migration rates, defined by Eq.~(\ref{mAndQDefn}), are all ${\mathcal O}(\theta)$ for some small parameter $\theta <<1$.  
There are two different time scales operating in this problem.  In terms of the diffusion time $t$, subsamples of the same allele type will coalesce 
to single ancestors within an island in ${\mathcal O}(1)$ time, whereas coalescence of the same allele type between islands or of different allele types 
within an island will take ${\mathcal O}(1/\theta)$ time.  Many authors have observed this behaviour, with early research by \citet{slatkin1981fixation,takahata1991genealogy,wakeley2001coalescent} and \citet{notohara2001structured}. 

A limit theorem where there are two time scales from a discrete model is Theorem 1 in \cite{mohle1998convergence}.  This is a useful theorem for models in 
population genetics. Let $\Pi_\alpha=A + B/\alpha + o(\alpha^{-1})$ as $\alpha \rightarrow \infty$ be a transition probability matrix such that 
$P= \lim_{m\to \infty}A^m$ exists.  
Then, with $G = PBP$,
\begin{equation}
\Pi(t') = \lim_{\alpha \to \infty}\Pi_\alpha ^{[\alpha t']} = Pe^{t'G},
\label{Mohle:0}
\end{equation}
are the transition functions of a continuous time process corresponding to a time $t'$ related to the discrete time $\tau$ by $\tau = [\alpha t']$.  
Usually $\alpha$ is proportional to the population size in a Wright-Fisher model.   An interpretation 
of $\Pi(t')$ is that in the limit, time scale transitions according to $P$ occur instantaneously compared to transitions with rates in $G$ which are not 
instantaneous.  We will not review this concept here and refer the reader to \cite{mohle1998convergence,mohle1998coalescent}.  

The theorem can be applied in the context of the model considered in this paper not by choosing $\alpha$ to be the population size, but by choosing 
$\alpha = M/\theta$, that is, the scale of the inverse migration and mutation rates, with the population size $M$ initially held fixed and $\theta \to 0$.  
Let $\Pi_{1/\theta}$ be the transition probability matrix in a Wright-Fisher model for changes 
in sample ancestry of a sample of $n$ genes taken from the $g$ islands.  This matrix has elements indexed by $(\bm{n}_1,\ldots,\bm{n}_g)$, 
$|\bm{n}_1| + \cdots + |\bm{n}_g| \leq n$ where $n_{ia}$, $1 \le i \le g$, $1 \leq a \leq K$ is the number of ancestral genes of allele type $a$ on island $i$. 
$\Pi_{1/\theta} = A + \theta B + \mathcal{O}(\theta^2)$ where $A$ is the transition probability matrix in a Wright-Fisher model where there is coalescence, 
but no migration or mutation. A precise form for the elements of $\Pi_{1/\theta}$ is difficult to find because of the general mutation structure where the 
probability of transitions forward in time is known, but backwards transition probabilities are not known. There will be an expansion to $\mathcal{O}(\theta)$ 
with a simple form for $A$, but not $B$.  Elements of $A^m$ converge to a matrix $P$ which has elements zero or one depending on the absorbing states 
where coalescence takes place between genes of the same type on the same island. That is
\begin{equation}
   P_{(\bm{n}_1,\ldots, \bm{n}_g), (\bm{l}_1,\ldots,\bm{l}_g)} =
    \prod_{i=1}^g\prod_{a=1}^K \delta_{l_{ia},\mathbb{I}\{n_{ia}>0\}}.
\end{equation}
M{\"o}hle's Theorem can now be applied for fixed population size $M$ as $\theta \to 0$ showing that (\ref{Mohle:0}) holds.
Now $B=\frac{1}{M}C_M$, where $C_M$ converges to $C$ as $M \to \infty$.  $P$ does not depend on $M$.
Therefore
    \begin{equation}
    \lim_{M\to \infty}\lim_{\theta \to 0}\Pi_{1/\theta}^{[t'\frac{M}{\theta}]}\to Pe^{t'G}
    \end{equation}
where $G=PCP$.
    
The approach taken in this paper is to investigate the stationary distribution in a diffusion process, or underlying dual coalescent process directly rather than use  
M{\"o}hle's theorem for convergence from the discrete Wright-Fisher model.

 \cite{wakeley2001coalescent} investigates an infinitely-many-demes model. In a pre-limit model there are $g$ demes, then as $g\to \infty$ in a time scale 
 proportional to $g$ generations there is an instantaneous scattering phase where individuals migrate to different demes, then a collecting phase where migration 
 of an individual to an occupied deme results in instantaneous coalescence. Mathematics used in describing this model is the convergence theorem with two 
 time scales of \cite{mohle1998convergence} where $\alpha=g$ in Eq. (\ref{Mohle:0}). \citet{vogl2003population} construct an MCMC algorithm for inference 
 in the infinitely-many-demes model which is appplied to real data.

 \citet{ethier1980diffusion} study convergence of Wright-Fisher models with two time scales to a diffusion model, with applications in population genetics. 
 \citet{wakeley2004many} use their results elegantly to study frequencies  in their infinitely-many-demes model, where time scales within demes are much faster 
 than the time scale across demes. A similar approach cannot be used for the finite number of demes in this paper, as the number of demes tending to infinity 
 is crucial in their approach.

\citet{gutenkunst2009inferring} have developed numerical software called $\partial a \partial i$ for determining the site frequency spectrum of a multiple-island 
Wright-Fisher diffusion with migrations and selection, with mutations modelled by setting boundary conditions at the corners $(x_1, x_2) = (0, 0)$ and $(1, 1)$ 
of the region $\Omega$. 

%
%
\section{Stationary distribution with slow mutation and migration rates}
\label{sec:StationaryDist}

We will demonstrate that in the limit of slow mutation and migration rates, an accurate approximation to the stationary solution of Eq.~(\ref{forwardKolmog2}) can be 
found as a set of effective line densities on the boundary of the region $\Omega$.  The method is similar to that used by 
\cite{burden2016approximate,burden2017rate} 
to find an approximate stationary solution to the multi-allele neutral Wright-Fisher model for an arbitrary instantaneous rate matrix.  

We begin by rescaling the mutation rates by a small parameter $\theta << 1$ via the equations 
\begin{equation}
\begin{split}
q_{12} = \theta \alpha, \qquad q_{21} = \theta \beta, \\
m_{12} = \theta \mu_{12}, \qquad m_{21} = \theta \mu_{21},  \label{alphaBetaMuDef}
\end{split}
\end{equation}
where $\alpha$, $\beta$, $\mu_{12}$ and $\mu_{21}$ are $O(1)$.  For instance, one might choose 
$\theta = \max(q_{12}, q_{21}, m_{12}, m_{21})$,
though this specific choice is not absolutely necessary.  
For notational convenience we will also set $x_1 = x$ ($=$ the relative abundance of allele $A_1$ on island-1) and $x_2 = y$ ($=$ the relative abundance 
of allele $A_1$ on island-2).  With this reparameterisation, and setting the time derivative to zero, Eq.~(\ref{forwardKolmog2}) becomes 
\begin{eqnarray}
0 & = & \frac{1}{2r_1} \frac{\partial^2}{\partial x^2} \left\{ x(1 - x) f(x,y) \right\}  
						      + \frac{1}{2r_2} \frac{\partial^2}{\partial y^2} \left\{ y(1 - y) f(x,y) \right\} \nonumber \\
	&  & - \theta\frac{\partial}{\partial x} \left[ \left\{ \mu_{12}(y - x) - \alpha x + \beta (1 - x) \right\} f(x,y) \right] \nonumber \\
	&  & - \theta\frac{\partial}{\partial y} \left[ \left\{ \mu_{21}(x - y) - \alpha y + \beta (1 - y) \right\} f(x,y) \right].  \nonumber \\ \label{forwardKolmogxy}
\end{eqnarray}
In the first instance our aim will be to find the stationary solution to leading order 
in $\theta$ as $\theta \rightarrow 0$ close to the boundary of $\Omega$.  

Consider first the region close to the edge $y = 0$.  Following the procedure described in Appendix~A of \cite{burden2016approximate}, without loss of generality 
we write the solution in the form 
\begin{equation}
f(x, y) = \theta^2 s(x) y^{\theta s(x) - 1} \sum_{k = 0}^\infty g_k(x) y^k.  \label{fAnsatz}
\end{equation}
This expansion is essentially a generalisation of the Frobenius method for solving ordinary differential equations~\citep{teschl2012ordinary} to our partial differential equation.  
The purpose of the function $s(x)$ is to capture the leading order power of $y$.  From numerical simulations, and from experience with the multi-allelic 
solution~\citep{burden2016approximate}, we will assume this exponent to be close to $-1$ for small $\theta$.   As we shall see, it will turn out that the choice of exponent and 
the overall normalisation $\theta^2$ ensure that $s(x)$ and $g_0(x)$ are analytic functions for $0 < x < 1$, which remain finite as $\theta \rightarrow 0$.  

\begin{table}[t]
\caption{Asymptotic behaviour of each term in Eq.~(\ref{forwardKolmogxy}). $\partial_x$ and $\partial_y$ mean $\partial/\partial x$ and $\partial/\partial y$
	respectively.}
\begin{center}
\begin{tabular}{l l l}
\hline
Term                                                & $\lim_{y \rightarrow 0} (\text{Term})$    & $\lim_{\theta \rightarrow 0}\int_0^\Lambda (\text{Term}) dy$ \\
\hline
$1/(2r_1) \partial_x^2 \{x(1 - x) f\}$ & $O(y^{\theta s(x) - 1} (\log y)^2 )$ &  $\frac{1}{2} \theta \partial_x^2 \{x(1 - x) g_0(x)\} + O(\theta^2)$  \\
$1/(2r_2) \partial_y^2 \{y(1 - y) f\}$ & $O(y^{\theta s(x) - 2})$                  &  Divergent \\
$-\theta \partial_x (\mu_{12} y f)$     & $O(y^{\theta s(x) } \log y )$ & $O(\theta^3)$\\
$-\theta \partial_x \{(- \mu_{12} x f - \alpha x + \beta(1 - x))f\} $  & $O(y^{\theta s(x) - 1} \log y )$ & $O(\theta^2)$\\
$-\theta \partial_y \{(\mu_{21} x + \beta) f\}$ & $O(y^{\theta s(x) - 2})$ & Divergent \\
$-\theta \partial_y \{(- \mu_{21} y - \alpha y - \beta y) f\}$ & $O(y^{\theta s(x) - 1})$ & $O(\theta^3)$\\
\hline
\end{tabular}
\end{center}
\label{fNearYEquals0Table}
\end{table}%

For fixed $x$ and $\theta$ the behaviour of each term in Eq.~(\ref{forwardKolmogxy}) as $y \rightarrow 0$ is as listed in the middle column of 
Table~\ref{fNearYEquals0Table}.  Keeping only the dominant $O(y^{\theta s(x) - 2})$ terms allows us to extract $s(x)$ as follows:  
\begin{eqnarray}
0 & = & \frac{\partial}{\partial y} \left\{\frac{1}{2r_2} \frac{\partial}{\partial y} (y f(x, y)) - \theta (\mu_{21} x + \beta)f(x, y) \right\} (1 + O(y)).   \nonumber \\ 
& = & \theta^2 s(x) g_0(x) \frac{\partial}{\partial y} \left\{\frac{1}{2r_2} \frac{\partial}{\partial y} y^{\theta s(x)} - \theta (\mu_{21} x + \beta)y^{\theta s(x) - 1} \right\} (1 + O(y)) \nonumber \\
& = & \theta^3 \frac{s(x) g_0(x) }{2r_2} [s(x) - 2 r_2(\mu_{21}x + \beta)] (\theta s(x) - 1) y^{\theta s(x) - 2} (1 + O(y)) \nonumber \\
																	\label{RHSforwardK}
\end{eqnarray}
This can only be achieved if 
\begin{equation}
s(x) = 2 r_2(\mu_{21}x + \beta).  \label{sOfX}
\end{equation}

Up to this point there is no requirement that $\theta$ should be small.  We now impose such a requirement, and note that as a consequence, Eqs.~(\ref{fAnsatz}) 
and (\ref{sOfX}) imply that $f(x, y)$ drops off rapidly away from the boundary at $y = 0$.  Now introduce a cutoff a distance $\Lambda$ from the boundary and 
define an effective line density $f_1^{\rm eff}(x)$, such that $f_1^{\rm eff}(x)\,dx$ is the probability contained in the region $[x, x + dx] \times [0, \Lambda]$.  Then 
\begin{eqnarray}
f_1^{\rm eff}(x) & = & \int_0^\Lambda f(x, y)\, dy \nonumber \\
& = & \theta^2 s(x) \sum_{k = 0}^\infty g_k(x) \int_0^\Lambda y^{\theta s(x) - 1 + k} dy \nonumber \\
& = & \theta g_0(x) \Lambda^{\theta s(x)} + \theta^2 \sum_{k=1}^\infty \frac{\Lambda^{\theta s(x) + k}}{\theta s(x) + k} \nonumber \\
& = & \theta g_0(x) + O(\theta^2), \qquad \mbox{as } \theta \rightarrow 0.  \label{IntOfF}
\end{eqnarray}
Importantly, we see that $\theta^{-1} f_1^{\rm eff}(x)$ is independent of $\Lambda$ in the absolute limit $\theta \rightarrow 0$, and that for practical purposes the approximation 
to a line density is accurate to leading order in $\theta$ provided 
\begin{equation}
\theta s(x) \left|\log\Lambda\right| << 1,		\label{LambdaRestriction}
\end{equation} 
or equivalently, $\Lambda >> e^{-1/\theta}$ when $s(x) = O(1)$.  Similarly we have that 
\begin{equation} 
\int_0^\Lambda yf(x, y)\, dy = O(\theta^2), \quad \int_0^\Lambda \partial_y [y f(x, y)]\, dy = O(\theta^2), \qquad \mbox{as } \theta \rightarrow 0. \label{otherIntTerms}
\end{equation} 
Thus the asymptotic behaviour of the integral of each term in Eq.~(\ref{RHSforwardK}) is as listed in the right-hand column of 
Table~\ref{fNearYEquals0Table}.  Note that by Eq.~(\ref{sOfX}) and a calculation similar to that leading to Eq.~(\ref{otherIntTerms}) the sum of the two integrals 
listed as divergent is also $O(\theta^2)$.  Integrating Eq.~(\ref{forwardKolmogxy}) term-by-term, dividing by $\theta$, and taking the limit $\theta \rightarrow 0$ then gives 
\begin{equation}
\frac{d^2}{dx^2} [x(1 - x)g_0(x)] = 0.  
\end{equation}
The general solution to this equation is 
\begin{equation}
g_0(x) = \frac{a + \phi}{x} + \frac{a - \phi}{1 - x}, 
\end{equation}
where $a$ and $\phi$ are arbitrary constants.  

\begin{figure}[t!]
\begin{center}
\centerline{\includegraphics[width=0.6\textwidth]{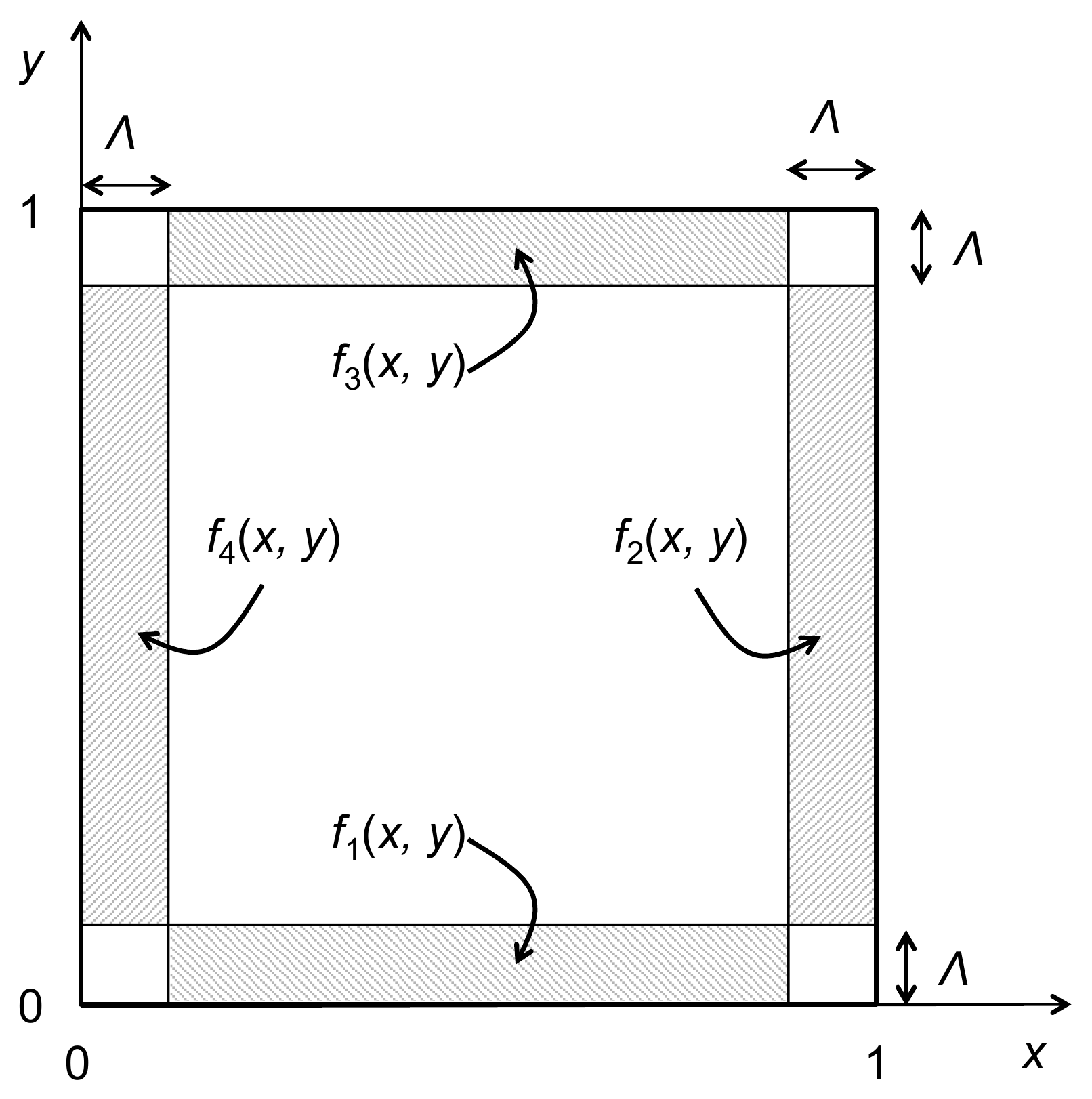}}
\caption{Regions in which the functions $f_1(x, y)$ to $f_4(x, y)$ defined in Eq.~(\ref{f1ToF4}) are valid.} 
\label{fig:UnitSquare}
\end{center}
\end{figure}

Repeating the above derivation on each of the 4 edges of the unit square we obtain, to lowest order in $\theta$, the following approximate solutions 
valid in the 4 regions shown in Fig.~\ref{fig:UnitSquare}:
\begin{equation}
\begin{split}
f_1(x, y) & = 2  \theta^2 r_1 r_2 (\mu_{21}x + \beta) y^{2 \theta r_2(\mu_{21} x + \beta) - 1} \left(\frac{A + \Phi}{x} + \frac{A - \Phi}{1 - x} \right), \\
f_2(x, y) & = 2  \theta^2 r_1 r_2 [\mu_{12}(1 - y) + \alpha] (1 - x)^{2 \theta r_1[\mu_{12}(1 - y) + \alpha] - 1} \left(\frac{B + \Phi}{y} + \frac{B - \Phi}{1 - y} \right), \\ 
f_3(x, y) & = 2  \theta^2 r_1 r_2 [\mu_{21}(1 - x) + \alpha] (1 - y)^{2 \theta r_2[\mu_{21}(1 - x) + \alpha] - 1} \left(\frac{C - \Phi}{x} + \frac{C + \Phi}{1 - x} \right), \\
f_4(x, y) & = 2  \theta^2 r_1 r_2 (\mu_{12}y + \beta) x^{2 \theta r_1(\mu_{12}y + \beta) - 1} \left(\frac{D - \Phi}{y} + \frac{D + \Phi}{1 - y} \right),
\end{split}									\label{f1ToF4}
\end{equation}
where the parameters $A$, $B$, $C$, $D$, and $\Phi$ are yet to be determined.  

Note that there is in general a net flux of probability flow around the boundary of the square, 
encapsulated in a single parameter $\Phi$.  To see that the same value of $\Phi$ is relevant to all 4 sides of $\Omega$,  
we temporarily reinstate the time derivative on the left hand side of Eq.~(\ref{forwardKolmogxy}) and integrate out from the boundary at $y = 0$ to the 
cutoff $\Lambda$ employed in Eq.~(\ref{IntOfF}) to obtain 
\begin{eqnarray}
\frac{\partial f_1^{\rm eff}}{\partial t} & = & \frac{1}{2r_1} \frac{\partial^2}{\partial x^2} \left\{ x(1 - x) f_1^{\rm eff} \right\} + \mbox{higher order in $\theta$} \nonumber \\
	& = & - \frac{\partial}{\partial x} \left[ - \frac{1}{2r_1} \frac{\partial}{\partial x} \left\{ x(1 - x) f_1^{\rm eff} \right\} \right] + \mbox{higher order in $\theta$}.
\end{eqnarray}
The factor in square brackets is the flux of probability from left to right across the bottom edge of the unit square.  Substituting in $f_1$ from Eq.~(\ref{f1ToF4}) 
gives 
\begin{eqnarray}
f_1^{\rm eff} (x) & = &  \theta r_1 \left( \int_0^\Lambda 2\theta r_2 (\mu_{21}x + \beta) y^{2 \theta r_2(\mu_{21} x + \beta) - 1} dy  \right)
                                                                  \left(\frac{A + \Phi}{x} + \frac{A - \Phi}{1 - x} \right)  \nonumber \\
                      & = &  \theta r_1 \left(\frac{A + \Phi}{x} + \frac{A - \Phi}{1 - x} \right)   + O(\theta^2),	\label{fEff1}
\end{eqnarray}
and hence to lowest order in $\theta$ the probability flux is 
\begin{equation}
- \frac{\theta}{2} \frac{d}{dx} \{(A + \Phi)(1 - x) + (A - \Phi)x \}  =   \theta \Phi.  \label{netFlux}
\end{equation}
The analogous calculation along the three remaining edges gives 
\begin{equation}
\begin{split}
f_2^{\rm eff} (y) &=  \theta r_2 \left(\frac{B + \Phi}{y} + \frac{B - \Phi}{1 - y} \right)   + O(\theta^2), \\
f_3^{\rm eff} (x) &=  \theta r_1 \left(\frac{C - \Phi}{x} + \frac{C + \Phi}{1 - x} \right)   + O(\theta^2), \\
f_4^{\rm eff} (y) &=  \theta r_2 \left(\frac{D - \Phi}{y} + \frac{D + \Phi}{1 - y} \right)   + O(\theta^2),
\end{split}			\label{fEff234}
\end{equation}
all of which lead to the same anticlockwise flux, as required if probability is to be conserved.  

The parameters $A$, $B$, $C$, $D$ and $\Phi$ are set by matching solutions at the corners.  For instance, for $(x, y)$ close to $(0, 0)$, 
Eq.~(\ref{f1ToF4}) implies that the dominant behaviour is 
\begin{equation}
f(x, y) = 2(A + \Phi)\theta^2 r_1 r_2 \beta y^{2\theta r_2\beta - 1} x^{-1} (1 + O(\theta)), \label{cornerFromF1}
\end{equation}
from $f_1$, and 
\begin{equation}
f(x, y) = 2(D -  \Phi)\theta^2 r_1 r_2 \beta x^{2\theta r_1\beta - 1} y^{-1} (1 + O(\theta)), \label{cornerFromF4} 
\end{equation}
from $f_4$.  These can only be consistent if $A + \Phi = D - \Phi$.  Applying a similar argument to the 4 corners gives 
\begin{equation}
\begin{split}
D - \Phi &= A + \Phi \\
(\mu_{21} + \beta)(A - \Phi) &= (\mu_{12} + \alpha)(B + \Phi) \\
B - \Phi &= C + \Phi \\
(\mu_{21} + \alpha)(C - \Phi) &= (\mu_{12} + \beta)(D + \Phi)
\end{split}									\label{AToDAndPhi}
\end{equation}
The solution of these equations, up to an overall factor $\kappa$, is 
\begin{equation}
\begin{split}
A &= \kappa(4\alpha + 3\mu_{12} + \mu_{21}), \\
B &= \kappa(4\beta + \mu_{12} + 3\mu_{21}), \\
C &= \kappa(4\beta + 3\mu_{12} + \mu_{21}), \\
D &= \kappa(4\alpha + \mu_{12} + 3\mu_{21}), \\
\Phi &= \kappa(\mu_{21} - \mu_{12}). 
\end{split} 	\label{ABCDSolution}
\end{equation}

The overall scale $\kappa$ is set by normalising the total probability to 1.  Noting that the contribution to the probability 
in the vicinity of the corners is $O(1)$, whereas the contribution in the vicinity of the edges (that is, the 4 shaded areas in 
Fig.~(\ref{fig:UnitSquare})) is $O(\theta)$, we observe that it is sufficient to consider only the corner contributions.  For instance, 
in the vicinity of the corner $(x, y) = (0, 0)$, in order to have an integrable singularity 
we expect $f(x, y) \sim \text{constant} \times x^{-1 + O(\theta)}y^{-1 + O(\theta)}$.  This is achieved in a way consistent with 
Eqs.~(\ref{cornerFromF1})) and (\ref{cornerFromF4}) if 
\begin{equation}
f(x, y) \sim 2 \theta^2 r_1 r_2 \beta (A + \Phi) x^{2\theta r_1 \beta - 1} y^{2\theta r_2 \beta - 1}.  
\end{equation}
The contribution from the corner $[0, \Lambda] \times [0, \Lambda]$ is then 
\begin{equation}
\int_0^\Lambda \int_0^\Lambda  f(x, y)  \,dx\,dy =  \frac{A + \Phi}{2\beta} + O(\theta) = \frac{\kappa}{\beta}(\mu_{12} + \mu_{21} + 2\alpha) + O(\theta).  
			\label{corner1}
\end{equation}

Suppose we define $P_{ab}$ to be the joint probability to zeroth order in $\theta$ that a single individual selected at random from island 1 is of allele type $A_a$, 
and a single individual selected at random from island 2 is of allele type $A_b$. These zero-order probabilities are precisely the corner probabilities.  
Since $x_1$ and $x_2$ are the relative frequencies of allele $A_1$ on islands 1 and 2 respectively, we have from Eq.~(\ref{corner1}) that 
\begin{equation}
P_{22} = \frac{\kappa}{\beta}(\mu_{12} + \mu_{21} + 2\alpha).  
\end{equation}
Similarly, calculating contributions from the corners $(x, y) = (1, 0)$, $(1, 1)$ and $(0, 1)$ gives respectively 
\begin{equation}
\begin{split}
P_{12} &=  \frac{B + \Phi}{2(\mu_{21} + \beta)} = 2\kappa,  	\\
P_{11} &=  \frac{C + \Phi}{2\alpha} = \frac{\kappa}{\alpha}(\mu_{12} + \mu_{21} + 2\alpha),  			\\
P_{21} &=  \frac{D + \Phi}{2(\mu_{21} + \alpha)} = 2\kappa.  	\\
\end{split}	\label{PijValues}
\end{equation}
To leading order in $\theta$ the sum of these must be 1, and hence 
\begin{equation}
\kappa = \frac{\alpha\beta}{(\alpha + \beta)[\mu_{12} + \mu_{21} + 2(\alpha + \beta)]}.  \label{kappaValue}
\end{equation}
The corner probabilities are then 
\begin{equation}
\left( \begin{array}{c}
P_{11}\\
P_{12}\\
P_{21}\\
P_{22}\ \end{array} \right) 
= \frac{1}{(\alpha + \beta)[\mu_{12} + \mu_{21} + 2(\alpha + \beta)]} \left( \begin{array}{c} 
(\mu_{12} + \mu_{21} + 2\beta)\beta \\
2\alpha\beta\\
2\alpha\beta\\
(\mu_{12} + \mu_{21} + 2\alpha)\alpha \end{array} \right).			\label{cornerProbs}
\end{equation}

As mentioned above, the choice of $\theta$ is to some extent arbitrary.  Equation~(\ref{alphaBetaMuDef}) implies that for given scaled rates $q_{ab}$ and $m_{ij}$,
 the numerical values of $\alpha$, $\beta$, 
$\mu_{12}$, and $\mu_{21}$ scale like $\theta^{-1}$, and Eqs.~(\ref{ABCDSolution}) and (\ref{kappaValue}) ensure that $A$, $B$, $C$, $D$ and $\Phi$ 
also scale like $\theta^{-1}$.  From Eqs.~(\ref{f1ToF4}), (\ref{fEff1}), (\ref{fEff234}) and (\ref{cornerProbs}) it is then clear that the approximate 
functions $f_1$ to $f_4$, $f_1^{\rm eff}$ to $f_4^{\rm eff}$ and probabilities $P_{ab}$ are independent of the initial choice 
of $\theta$ for given $q_{ab}$ and $m_{ij}$.  

More specifically, in terms of the scaled rates, the principal results are summarised to leading order in $\theta$ as: from Eq~(\ref{f1ToF4}), 
\begin{equation}
\begin{split}
f_1(x, y) & = 8 \kappa r_1 r_2 (m_{21}x + q_{21}) y^{2 r_2(m_{21} x + q_{21}) - 1} \left(\frac{q_{12} + \frac{1}{2}m}{x} + \frac{q_{12} + m_{12}}{1 - x} \right), \\
f_2(x, y) & = 8 \kappa r_1 r_2 [m_{12}(1 - y) + q_{12}] (1 - x)^{2 r_1[m_{12}(1 - y) + q_{12}] - 1} \left(\frac{q_{21} + m_{21}}{y} + \frac{q_{21} + \frac{1}{2}m}{1 - y} \right), \\ 
f_3(x, y) & = 8 \kappa r_1 r_2 [m_{21}(1 - x) + q_{12}] (1 - y)^{2 r_2[m_{21}(1 - x) + q_{12}] - 1} \left(\frac{q_{21} + m_{12}}{x} + \frac{q_{21} + \frac{1}{2}m}{1 - x} \right), \\
f_4(x, y) & = 8 \kappa r_1 r_2 (m_{12}y + q_{21}) x^{2 r_1(m_{12} y + q_{21}) - 1} \left(\frac{q_{12} + \frac{1}{2}m}{y} + \frac{q_{12} + m_{21}}{1 - y} \right);
\end{split}									\label{f1ToF4ScaledRates}
\end{equation}
from Eqs.~(\ref{fEff1}) and (\ref{fEff234}) for the effective line densities, 
\begin{equation}
\begin{split}
f_1^{\rm eff} (x) &=  4 \kappa r_1 \left(\frac{q_{12} + \frac{1}{2}m}{x} + \frac{q_{12} + m_{12}}{1 - x} \right),	\\
f_2^{\rm eff} (y) &=  4 \kappa r_2 \left(\frac{q_{21} + m_{21}}{y} + \frac{q_{21} + \frac{1}{2}m}{1 - y} \right), \\
f_3^{\rm eff} (x) &=  4 \kappa r_1 \left(\frac{q_{21} + m_{12}}{x} + \frac{q_{21} + \frac{1}{2}m}{1 - x} \right), \\
f_4^{\rm eff} (y) &=  4 \kappa r_2 \left(\frac{q_{12} + \frac{1}{2}m}{y} + \frac{q_{12} + m_{21}}{1 - y} \right);
\end{split}			\label{fEffScaledRates}
\end{equation}
and from Eq.~(\ref{cornerProbs}) for the corner probabilities, 
\begin{equation}
\left( \begin{array}{c}
P_{11}\\
P_{12}\\
P_{21}\\
P_{22}\ \end{array} \right) 
= \frac{1}{q(m + 2q)} \left( \begin{array}{c} 
(m + 2q_{21})q_{21} \\
2q_{12} q_{21}\\
2q_{12} q_{21}\\
(m + 2q_{12})q_{12} \end{array} \right),			\label{cornerProbsScaledRates}
\end{equation}
where
\begin{equation}
\kappa = \frac{q_{12} q_{21}}{q(m + 2q)}, 
\qquad
q = q_{12} + q_{21}, \qquad m = m_{12} + m_{21},    \label{kappaMQDefns}
\end{equation}
and $r_1$ and $r_2$ are defined by Eq.~(\ref{relPopSizes}).  

Note that the asymmetry in the effective line densities, Eq.~(\ref{fEffScaledRates}) is due to a net flux of probability anticlockwise around the boundary 
of $\Omega$ when $m_{12} \ne m_{21}$ (see Eqs.~(\ref{netFlux}) and (\ref{ABCDSolution})). 
In this sense the line density differs from the small-$\theta$ limit of Wright's well known beta-function solution to the 2-allele neutral diffusion, for which 
the boundary conditions at $x = 0$ and $1$ constrain the flux to be zero. 
%
%
\section{Alternative derivation of $P_{ab}$ and $f_1^{\rm eff} (x)$: Backward generator}
\label{sec:AlternativeFromBackward}

In Section~\ref{sec:StationaryDist} the probabilities $P_{ab}$ are defined to be the joint probability that a single individual selected at 
random from island 1 is of allele type $A_a$, and a single individual selected at random from island 2 is of allele type $A_b$.  These probabilities can be 
written to zeroth order in $\theta$ as 
\begin{equation}
\begin{split}
P_{11} &= \mathbb{E}_0\big [X_1X_2 \big],	\\
P_{12} &= \mathbb{E}_0\big [X_1(1 - X_2) \big], \\
P_{21} &= \mathbb{E}_0\big [(1 - X_1)X_2 \big],	\\
P_{22} &= \mathbb{E}_0\big [(1 - X_1)(1 - X_2)\big],	\label{PabAsExpectations}
\end{split}
\end{equation}
where $X_i = X_i(\infty)$ are the type-$A_1$ allele frequencies defined in Eq.~(\ref{tAndXDefn}) in the stationary distribution.  

Here we give an alternative derivation of the analytic formulae for these probabilities to zeroth order in $\theta$, namely Eq.~(\ref{cornerProbsScaledRates}).  
The starting point is a general result \citep[][Section~3.6]{etheridge2011some} that 
\begin{equation}
\mathbb{E}\big [ {\cal L} g(\mathbf X) \big] = 0, \label{ExpLG}
\end{equation}
for any function $g$ in the domain of the generator $\cal L$ of the backward Kolmogorov equation, with expectation in the stationary distribution of the  
process if it exists.  For the 2-island, 2-allele model the backward generator is~\citep{de2004importance}
\begin{eqnarray}
{\cal L} &= &\frac{1}{2}\sum_{i=1}^2\frac{1}{r_i}x_i(1-x_i)\frac{\partial^2}{\partial x_i^2}
\nonumber \\
&&~~+\Big (m_{12}(x_2-x_1)-x_1q_{12}+(1-x_1)q_{21}\Big )\frac{\partial}{\partial x_1}
\nonumber \\
&&~~+ \Big (m_{21}(x_1-x_2)-x_2q_{12}+(1-x_2)q_{21}\Big )\frac{\partial}{\partial x_2}.
\label{gen:0}
\end{eqnarray}

As before we assume the migration rates $m_{ij}$ and mutation rates $q_{ab}$ to be $O(\theta)$ as $\theta \rightarrow 0$.  
For the remainder of this section, 
let $\mathbb{E}_0$ denote expectation to order zero in $\theta$ and $\mathbb{E}_1$ expectation up to first order in $\theta$ (including zero order terms).  
It is possible to calculate the probabilities in Eq.~(\ref{PabAsExpectations}) exactly, but with small rates this is easier. 
Then, 
\begin{equation}
\mathbb{E}\big [X_1\big ]=
\mathbb{E}_0\big [X_1^2\big ]=
\mathbb{E}\big [X_2\big ]=
\mathbb{E}_0\big [X_2^2\big ]=\pi_1,
\label{simple:0}
\end{equation}
where
\begin{equation}
(\pi_1, \pi_2) = \frac{1}{q_{12} + q_{21}} (q_{21}, q_{12})		\label{piStationary}
\end{equation}
is the stationary left-eigenvector of the mutation rate matrix.  To confirm Eq.~(\ref{simple:0}), note that  
\begin{eqnarray}
0&=&\mathbb{E}\big [{\cal L}(X_1)\big ]
\nonumber \\
&=& m_{12}\mathbb{E}\big [X_2-X_1\big ] - \mathbb{E}\big [X_1]q_{12} + q_{21}\mathbb{E}\big [1-X_1\big],	
\nonumber \\
0&=&\mathbb{E}\big [{\cal L}(X_2)\big ]
\nonumber \\
&=& m_{21}\mathbb{E}\big [X_1-X_2\big ] - \mathbb{E}\big [X_2]q_{12} + q_{21}\mathbb{E}\big [1-X_1\big].	
\end{eqnarray}
The exact solution to these equations is $\mathbb{E}\big [X_1\big] =\mathbb{E}\big [X_2\big]=\pi_1$.  Then, 
\begin{eqnarray}
0 &=&	\mathbb{E}\big [{\cal L}(X_1^2)\big ]
\nonumber \\
&=& r_1^{-1}\mathbb{E}\big [X_1(1-X_1)\big ]
+ 2 \mathbb{E}\big [X_1\big (m_{12}(X_2-X_1) -X_1q_{12}+(1-X_1)q_{21}\big )\big ]
\nonumber \\
&=& r_1^{-1}\Big (\mathbb{E}\big [X_1] - \mathbb{E}\big [ X_1^2] \Big ) + \text{small~order~terms}.
\end{eqnarray}
Therefore $\mathbb{E}_0\big [X_1^2\big ] = \mathbb{E}\big [X_1\big ]=\pi_1$, and similarly for $X_2$.

Applying Eq.~(\ref{ExpLG}) with $g(\mathbf x) = x_1 x_2$, we have 
\begin{eqnarray}
0 &=& \mathbb{E}\Big [{\cal L}(X_1X_2)\Big ]	
\nonumber \\
&=& m_{12}\mathbb{E}\big [X_2^2\big ] - m_{12}\mathbb{E}\big [X_1X_2\big ] - q_{12}\mathbb{E}\big [X_1X_2\big ] + q_{21}\mathbb{E}\big [X_2(1-X_1)\big ]
\nonumber \\
&&~+m_{21}\mathbb{E}\big [X_1^2\big ] - m_{21}\mathbb{E}\big [X_1X_2\big ]
-q_{12}\mathbb{E}\big [X_1X_2\big ] + q_{21}\mathbb{E}\big [X_1(1-X_2)\big ]
\nonumber \\
&=&-(m_{12}+m_{21}+2(q_{21}+q_{12}))\mathbb{E}\big [X_1X_2\big ]+ (m_{12}+m_{21}+2q_{21})\pi_1 
\nonumber\\
& & \qquad + \text{ higher order terms.}
\label{solution:0}
\end{eqnarray}
The zero order approximation (\ref{simple:0}) is used to obtain the last line.
Therefore to  order zero we recover Eq.~(\ref{cornerProbsScaledRates}),
\begin{equation}
P_{11} = \mathbb{E}_0\big [X_1X_2]  = \frac{m_{12}+m_{21}+	2q_{21}}
{m_{12}+m_{21}+2(q_{21}+q_{12})}\pi_1 = \frac{(m + 2q_{21})q_{21}}{(m + 2q)q},
\label{a:0}
\end{equation}
where $m$ and $q$ are defined by Eq.~(\ref{kappaMQDefns}).  By symmetry
\begin{equation}
P_{22} = \mathbb{E}_0\big [(1-X_1)(1-X_2)]  = \frac{(m + 2q_{12})q_{12}}{(m + 2q)q}.
\label{a:1}
\end{equation}
Now
\begin{equation}
P_{12} = \mathbb{E}_0\big [X_1(1-X_2)\big ] = \pi_1-\mathbb{E}_0\big [X_1X_2\big ]
= \frac{2q_{12} q_{21}}{(m + 2q)q},	
\label{a:2}
\end{equation}
and by symmetry 
\begin{equation}
P_{21} = \mathbb{E}_0\big [(1-X_1)X_2]  = \frac{2q_{12} q_{21}}{(m + 2q)q} = P_{12}.  	
\label{a:3}
\end{equation}

It is easy to confirm that the 4 probabilities in Eqs.~(\ref{a:0}) to (\ref{a:3}) sum to 1.  Note too that if $m_{12}=m_{21}=0$ then the four probabilities are found 
from the product measure, respectively $\pi_1^2$, $\pi_2^2$, $\pi_1\pi_2$, $\pi_2\pi_1$,  as they should be.

It is also possible to calculate to $\mathcal{O}(\theta)$ the marginal probability that subsamples are monomorphic, $\mathbb{E}\big [X_1^n\big ]$ and $\mathbb{E}\big [X_2^n\big ]$. This will give an indication of the order in $n$ for which the approximations hold.
Consider
\begin{equation}
{\cal L}x_1^n = \frac{n(n-1)}{2}\cdot \frac{1}{r_1}x_1^{n-1}(1-x_1) + 
n \big (m_{12}(x_2-x_1) - x_1q_{12} + (1-x_1)q_{21}\big )x_1^{n-1}.
\end{equation}
Taking expectations and setting $\mathbb{E}\big [{\cal L}X_1^n\big ] = 0$,
\begin{eqnarray}
\lefteqn{\mathbb{E}\big [X_1^n\big ] - \mathbb{E}\big [X_1^{n-1}\big ]} \nonumber \\
& = &\frac{2r_1}{n-1}\mathbb{E}\big [ \big (m_{12}\big (1-X_1 - (1-X_2)\big )) - X_1q_{12} + (1-X_1)q_{21}\big )X_1^{n-1}\big ]. \nonumber \\
\end{eqnarray}
Taking into account the order of the terms
\begin{eqnarray*}
\mathbb{E}_1\big [X_1^n\big ] - \mathbb{E}_1\big [X_1^{n-1}\big ]
 &=& - \frac{2r_1}{n-1}\mathbb{E}_0\big [\big (m_{12}(1-X_2) + X_1q_{12}\big )X_1^{n-1}\big ]
\nonumber \\
&=&- \frac{2r_1}{n-1}\big (m_{12}P_{12} + \pi_1q_{12}\big ).
\end{eqnarray*}
Thus
\begin{equation}
\mathbb{E}_1\big [X_1^n\big ] = \pi_1 - 2r_1 \big (m_{12}P_{12} + \pi_1q_{12}\big ) \sum_{j=1}^{n-1}\frac{1}{j}
\end{equation}
and similarly
\begin{equation}
\mathbb{E}_1\big [X_2^n\big ] = \pi_1 - 2r_2 \big (m_{21}P_{21} + \pi_1q_{12}\big )\sum_{j=1}^{n-1}\frac{1}{j}.
\end{equation}
These estimates are $\mathcal{O}(\theta)$, however to hold for larger $n$ and remain non-negative
\begin{equation}
2r_1\Big (\frac{m_{12}P_{12}}{\pi_1}+q_{12}\Big ) \log n \ll 1, \>
2r_2\Big (\frac{m_{21}P_{21}}{\pi_1}+q_{12}\Big ) \log n \ll 1.
\end{equation}

We next calculate the effective line density $f_1^{\rm eff} (x)$, where $X_2=0$ and $X_1$ varies.
We first calculate $\mathbb{E}_1\big [X_1^{n_1}(1-X_1)^{n_2}(1-X_2)\big ]$ where $n_1, n_2 \geq 1$, and from this deduce the line density.
In calculating expectations $\mathbb{E}_0$ acting on powers of $X_1,X_2$ we are effectively just using the corner probabilities where $X_1$ and $X_2$ 
can only take values $0$ or $1$. Thus
\begin{equation}
\mathbb{E}_0\big [X_1^n(1-X_2)\big ]
= P_{12},\quad n\geq 1.
\end{equation}
We will also make use of the zero-order expectations 
\begin{equation}
\begin{split}
&\mathbb{E}_0\big [X_1^n \big ] = \mathbb{E}_0\big [X_1 \big ] = \pi_1, \\ 
&\mathbb{E}_0\big [X_1^n(1 - X_2)X_2 \big ] = 0, \\ 
&\mathbb{E}_0\big [X_1(1 - X_1) \big ] = 0. 
\end{split}
\end{equation}
Now consider a recursive calculation of $\mathbb{E}_1\big [X_1^n(1-X_2)\big ]$ from the stationary equation $0=\mathbb{E}\big [{\cal L}X_1^n(1-X_2)\big ]$.  
For $n > 1$, 
\begin{eqnarray}
0&=&\frac{1}{2r_1}n(n-1)
\Big (\mathbb{E}_1\big [X_1^{n-1}(1-X_2)\big ] - \mathbb{E}_1\big [X_1^{n}(1-X_2)\big ] \Big )
\nonumber \\
&&~+n\mathbb{E}_0\big [X_1^{n-1}(1-X_2)
\big (m_{12}(X_2-X_1) - X_1q_{12} + (1-X_1)q_{21}\big )\big ]
\nonumber \\
&&~- \mathbb{E}_0\big [X_1^n\big (m_{21}(X_1-X_2) - X_2q_{12} + (1-X_2)q_{21}\big )\big ]
\nonumber \\
&=&\frac{1}{2r_1}n(n-1)
\Big (\mathbb{E}_1\big [X_1^{n-1}(1-X_2)\big ] - \mathbb{E}_1\big [X_1^{n}(1-X_2)\big ] \Big )
\nonumber \\ 
&&-nP_{12}\Big (m_{12} + q_{12}\Big )
- \Big (P_{12}m_{21}+q_{12}P_{12} - q_{12}\pi_1 + q_{21}P_{12}\Big ).
\end{eqnarray} 
This gives the recursive rule
\begin{equation}
\mathbb{E}_1\big [X_1^{n}(1-X_2)\big ]  = \mathbb{E}_1\big [X_1^{n-1}(1-X_2)\big ] - \frac{c_1}{n - 1} - \frac{c_2}{n(n - 1)}, 
\end{equation}
where 
\begin{equation}
\begin{split}
c_1 &= 2r_1P_{12}\Big (m_{12}+q_{12}\Big ), \\
c_2 &= 2r_1\Big (P_{12}m_{21}+q_{12}P_{12} - q_{12}\pi_1 + q_{21}P_{12}\Big ).
\end{split}
\end{equation}
The recursion gives  
\begin{eqnarray}
\mathbb{E}_1\big [X_1^n(1-X_2)\big ] &=& 
\mathbb{E}_1\big [X_1(1-X_2)\big ] -c_1\sum_{j=2}^n\frac{1}{j-1} - c_2\sum_{j=2}^n \frac{1}{j(j-1)}
\nonumber \\
&=&\mathbb{E}_1\big [X_1(1-X_2)\big ] - c_1\sum_{j=1}^{n-1}\int_0^1x^{j-1}dx 
- c_2\left(1-\frac{1}{n}\right)
\nonumber \\
&=& \mathbb{E}_1\big [X_1(1-X_2)\big ] - c_2 - c_1\int_0^1\frac{1-x^{n-1}}{1-x}dx
+ c_2\int_0^1x^{n-1}dx. \nonumber \\
\label{calculation:0}
\end{eqnarray}
Now it is possible to calculate $\mathbb{E}_1\big [X_1^{n_1}(1-X_1)^{n_2}(1-X_2)\big ]$, which is $O(\theta)$ for $n_1, n_2\geq 1$. 
Expanding $(1-X_1)^{n_2}$ and substituting from (\ref{calculation:0}):
\begin{eqnarray}
\lefteqn{\mathbb{E}_1\big [X_1^{n_1}(1-X_1)^{n_2}(1-X_2)\big ]} \nonumber \\ 
&=& \sum_{k=0}^{n_2}{{n_2}\choose k}(-1)^k\mathbb{E}_1\big [X_1^{{n_1}+k}(1-X_2)\big ]
\nonumber \\
&=& \sum_{k=0}^{n_2}{{n_2}\choose k}(-1)^k\left(
-c_1\int_0^1\frac{1-x^{{n_1}+k-1}}{1-x}dx + c_2\int_0^1x^{{n_1}+k-1}dx\right)
\nonumber \\
&=& c_1 \int_0^1x^{{n_1}-1}(1-x)^{{n_2}-1}dx + c_2 \int_0^1x^{{n_1}-1}(1-x)^{n_2}dx.
\label{calculation:1}
\end{eqnarray}
Note that the terms  $\mathbb{E}_1\big [X_1(1-X_2)\big ] - c_2$ do not contribute 
because $\sum_{k=0}^{n_2}{{n_2}\choose k}(-1)^k = (1 - 1)^{n_2} = 0$.
By writing this expectation as 
\begin{equation}
\mathbb{E}_1\big [X_1^{n_1}(1-X_1)^{n_2}(1-X_2)\big ] = \int_0^1 x^{n_1}(1-x)^{n_2} f_1^{\rm eff} (x) dx, 
\end{equation}
we read off the line density
\begin{eqnarray}
f_1^{\text{eff}}(x) &=& \frac{c_1}{x(1-x)} + \frac{c_2}{x}
\nonumber \\
&=& \frac{c_1+c_2}{x} + \frac{c_1}{1 - x}.
\end{eqnarray}
Evaluating the constants with the aid of Eqs.~(\ref{kappaMQDefns}), ({\ref{piStationary}) and (\ref{a:2}) gives 
\begin{equation}
\begin{split}
c_1 &= 4\kappa r_1 (q_{12} + m_{12}), \\
c_1 + c_2 &= 4\kappa r_1 (q_{12} + \tfrac{1}{2}m),
\end{split}
\end{equation}
in agreement with Eq.~(\ref{fEffScaledRates}).  The remaining three line densities are found in a similar fashion.  

%
%
\section{Alternative derivation of  the probabilities $P_{ab}$: The coalescent}
\label{sec:AlternativeFromCoalescent}

The probability $P_{ab}$ that one individual chosen at random from island-1 and one individual chosen at random from island-2 are of allele types 
$A_a$ and $A_b$ respectively can also be derived, to lowest order in $\theta$, using the coalescent.  

Consider the most recent common ancestor of two individuals, one chosen from each island.  
Tracing lines of descent backwards in time, it is clear that at least one migration event has occurred in the common ancestry of these individuals since the common ancestor.  
The random time $T_m$ since the most recent migration event is an exponential random variable with rate 
\begin{equation}
m = m_{12} + m_{21}.  
\end{equation}
Since $m$ is of order $\theta$, we have that $T_m = O(\theta^{-1})$.  Immediately before the migration event, both lines of descent inhabit the same island, and so 
standard coalescent theory tells us that the time elapsed between the common ancestor and the migration event is $O(1)$.  
It follows that, since the mutation rates $q_{ab}$ are of order $\theta$, to 
calculate the probability $P_{ab}$ to lowest order in $\theta$, it is sufficient firstly to 
consider only the time since the migration event, and secondly to assume that the ancestors at the time of the migration event were of the same allele type.  

Thus to $O(1)$ in $\theta$, we have two independently mutating parallel lines of descent, both descending from the same allele type.  Suppose we first fix the time 
since the migration event to be $t$.  For any $2 \times 2$ rate matrix $Q$ it is well known that 
\begin{equation}
e^{tQ} = I + \frac{1}{q} \left( 1 - e^{-tq} \right) Q, 	\label{expTQ}
\end{equation}
where $I$ is the identity matrix and $q = q_{12} + q_{21}$.  
Then from Eqs.~(\ref{piStationary}) and (\ref{expTQ}), 
\begin{eqnarray}
P_{12}(\text{fixed time }t) & = & \pi_1 (e^{tQ})_{11} (e^{tQ})_{12} + \pi_2 (e^{tQ})_{21} (e^{tQ})_{22}	\nonumber \\
	& = & \pi_1 \left[ 1 - \frac{q_{12}}{q} (1 - e^{-tq}) \right] \frac{q_{12}}{q} (1 - e^{-tq})   \nonumber \\
	 & & +         \pi_2  \frac{q_{21}}{q} (1 - e^{-tq})  \left[ 1 - \frac{q_{21}}{q} (1 - e^{-tq}) \right]  \nonumber \\
	& = & \frac{q_{12}}{q} (1 - e^{-2tq}) \pi_1.  
\end{eqnarray}
Taking the expectation with respect to the random time $T_m$ then gives 
\begin{eqnarray}
P_{12} & = & \int_0^\infty \frac{q_{12}}{q} (1 - e^{-2tq}) \pi_1 \times me^{-mt} dt \nonumber \\
	&= & \frac{q_{12}\pi_1}{q} \int_0^\infty (1 - e^{-2qx/m}) e^{-x} dx.  
\end{eqnarray}
The change of variable in the last line illustrates the point that it is not sufficient to consider only a single mutation event along each descendent line,   
even though our aim is only to calculate $P_{12}$ to leading order: The small parameters occur in the exponential in the ratio $q/m$, which is $O(1)$ as $\theta \rightarrow 0$.     
Evaluation of the integral and some straightforward rearranging gives 
\begin{equation}
P_{12} = \frac{2q_{12}}{m + 2q} \pi_1,
\end{equation} 
which agrees with Eq.~(\ref{cornerProbsScaledRates}).  

Similarly 
\begin{equation}
P_{11}(\text{fixed time }t) = \left[ 1 - \frac{q_{12}}{q} (1 - e^{-2tq}) \right]  \pi_1, 
\end{equation}
and hence, after taking the expectation with respect to $T_m$ and rearranging, yields 
\begin{equation}
P_{11} = \frac{m + 2q_{21}}{m + 2q} \pi_1, 
\end{equation} 
also agreeing with Eq.~(\ref{cornerProbsScaledRates}).  The probabilities $P_{21}$ and $P_{22}$ follow by symmetry.  

%
%
\section{Comparison with the numerical simulation}
\label{sec:ComparisonNumerical}

Below we compare numerical stationary distributions of the transition matrix Eq.~(\ref{transitionMatrix}) with our 
approximate analytic solutions to the diffusion limit forward Kolmogorov equation near the boundary of $\Omega$, namely Eq.(\ref{f1ToF4ScaledRates}), 
and with the approximate line densities on the boundary of $\Omega$, Eq.~(\ref{fEffScaledRates}).  

\begin{figure}[t!]
\begin{center}
\centerline{\includegraphics[width=\textwidth]{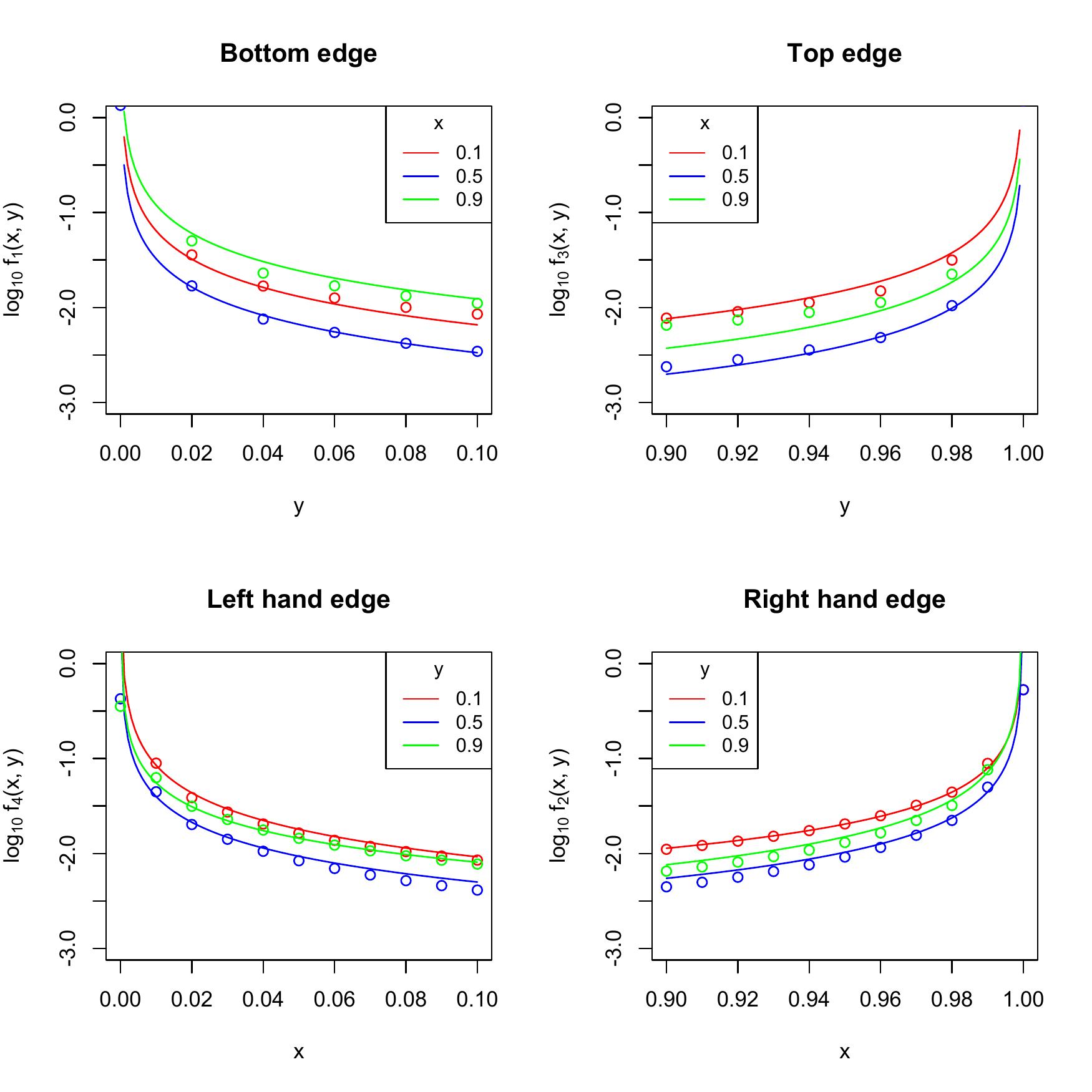}}
\caption{Logarithmic plots of the approximate analytic solutions $f_1$ to $f_4$ given in Eq.~(\ref{f1ToF4ScaledRates}) along traverse lines perpendicular to 
and in the vicinity of the edges of the unit square $\Omega = [0, 1] \times [0, 1]$.  Also plotted (circles) is the numerically determined solution 
obtained as $M_1 M_2$ times the stationary eigenvector of the transition matrix Eq.~(\ref{transitionMatrix}).  Parameters for the approximate analytic solutions 
are given in Eq.~(\ref{numericScaledParametersA}) and for the numerical simulation are as in the caption to Fig.~\ref{fig:StationaryPersp}(a) and (b).} 
\label{fig:TraversePlotsA}
\end{center}
\end{figure}

Figure~\ref{fig:TraversePlotsA} shows plots of the approximate analytic solutions $f_1$ to $f_4$ given in Eq.~(\ref{f1ToF4ScaledRates}) along traverse lines perpendicular to 
and in the vicinity of the edges of the unit square $\Omega = [0, 1] \times [0, 1]$ for the scaled parameters in Eq.~(\ref{numericScaledParametersA}), 
together with appropriately scaled probabilities of the numerically determined stationary solution shown in Figs.~\ref{fig:StationaryPersp} (a) and (b).  
The analytic solutions are a close match, but begin to drift off slightly as one moves away from the boundary.  

Figure~\ref{fig:TraversePlotsB} shows the analogous plot to Fig.~\ref{fig:TraversePlotsA}, except the that both mutation and migration rates are greater by a factor of 10, 
that is, 
\begin{equation}
\begin{split}
q_{12} = 0.75 \times 10^{-1}, \qquad q_{21} = 1.5 \times 10^{-1}\\
m_{12} = 12 \times 10^{-1}, \qquad m_{21} = 0.3 \times 10^{-1}.  
\end{split}		\label{numericScaledParametersB}
\end{equation}
In this case the approximate functions $f_1$ to $f_4$ are clearly a poor approximation.  In general, we find that Eq.~(\ref{f1ToF4ScaledRates}) begins to fail  
if $\theta = \max(q_{12}, q_{21}, m_{12}, m_{21})$ exceeds $0.1$.  

\begin{figure}[t!]
\begin{center}
\centerline{\includegraphics[width=\textwidth]{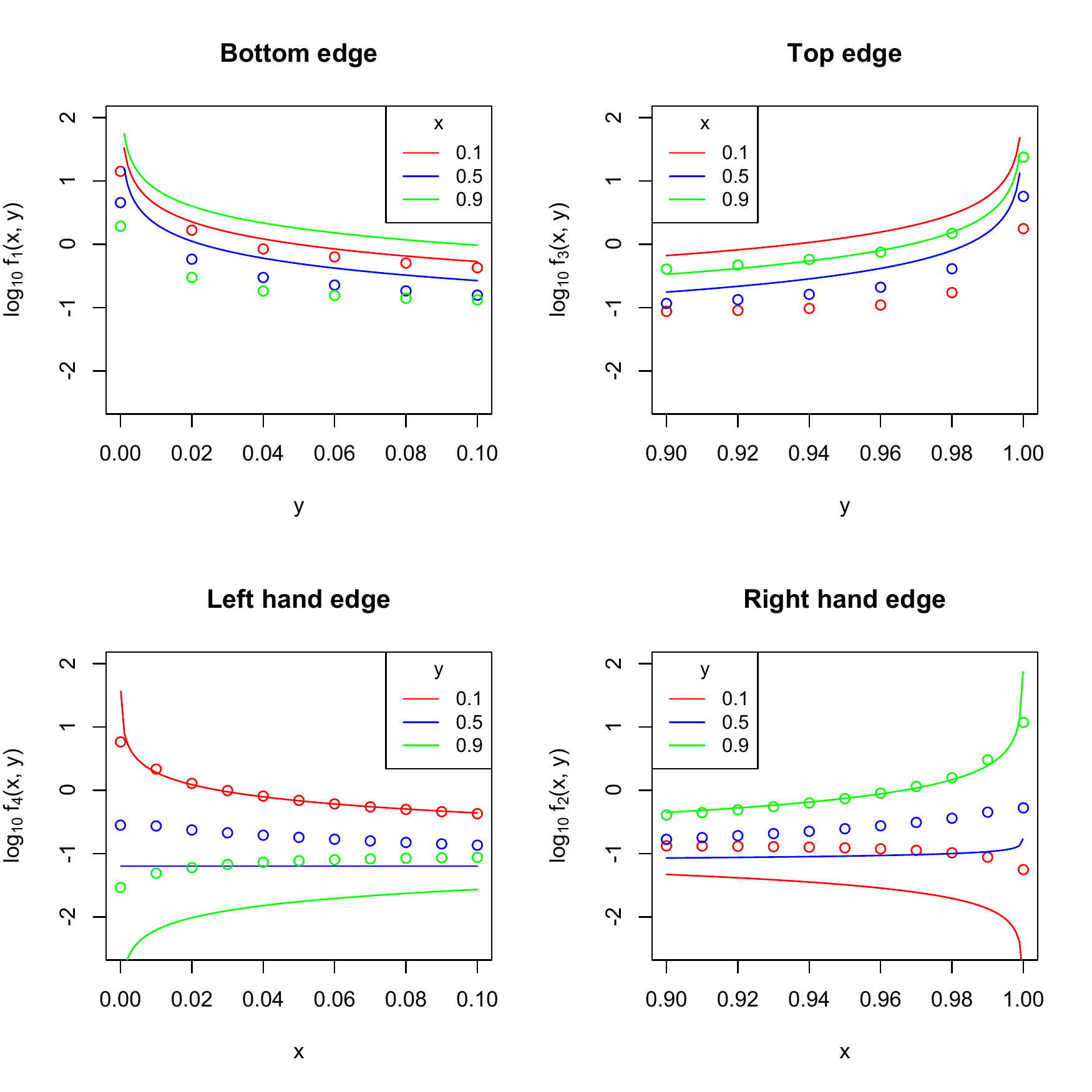}}
\caption{The same as Fig.~\ref{fig:TraversePlotsA}, except that mutation rates have been multiplied by a factor of 10.  That is, scaled rates are as in 
Eq.(\ref{numericScaledParametersB}), and the unscaled rates are $u_{12} = 0.5 \times 10^{-3}$, $u_{21} = 1 \times 
10^{-3}$, $v_{12} = 8 \times 10^{-3}$, $v_{21} = 0.2 \times 10^{-3}$
} 
\label{fig:TraversePlotsB}
\end{center}
\end{figure}

\begin{figure}[t!]
\begin{center}
\centerline{\includegraphics[width=\textwidth]{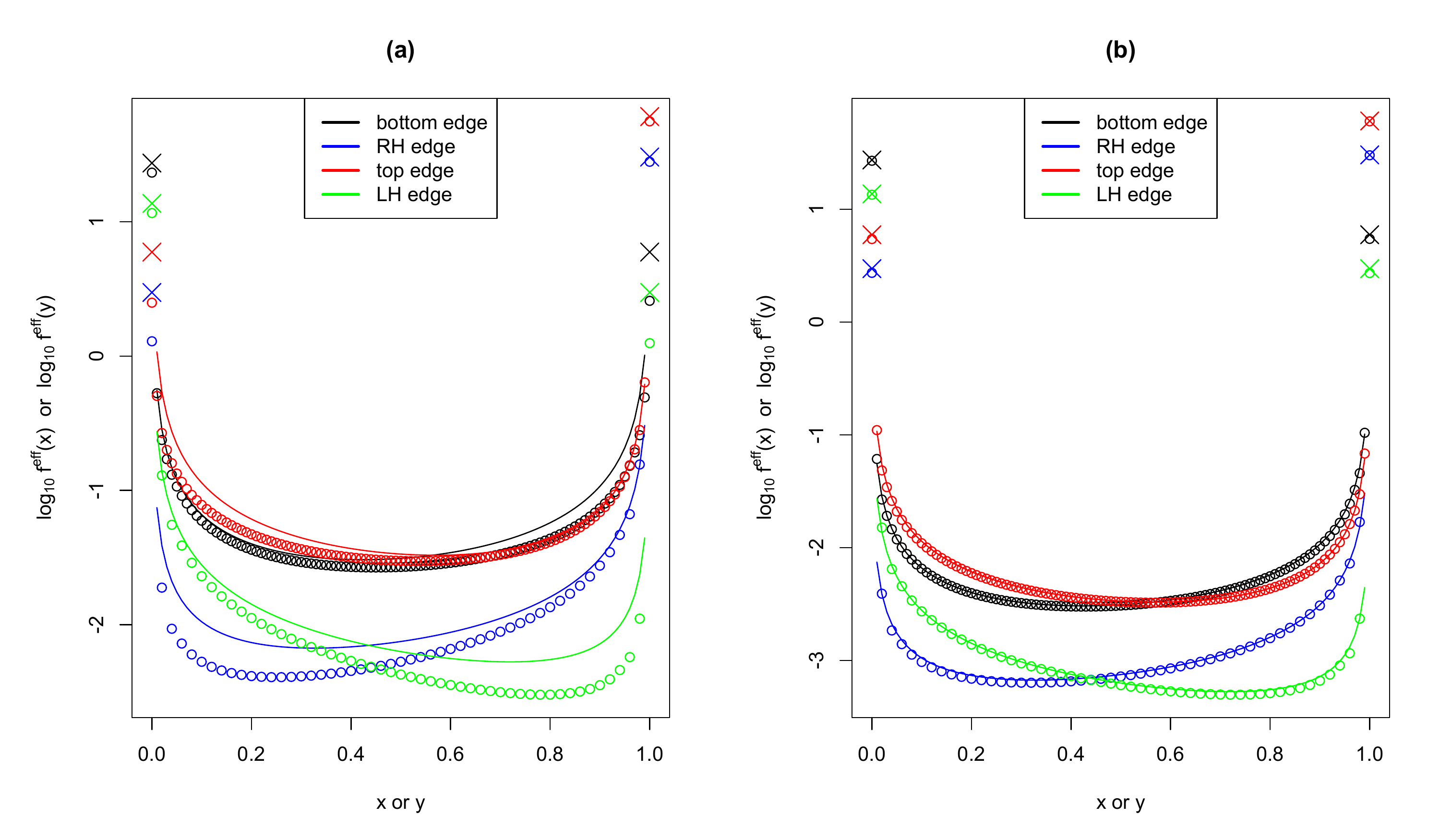}}
\caption{Logarithmic plots of the approximate effective line densities $f_1^{\rm eff}$ to $f_4^{\rm eff}$ given in Eq.~(\ref{fEffScaledRates}) along 
the 4 edges of the unit square $\Omega = [0, 1] \times [0, 1]$.  Also plotted as circles is the numerically determined stationary eigenvector of the transition matrix 
Eq.~(\ref{transitionMatrix}) along the 4 edges of the square lattice $\{0, \ldots, M_1\}\times\{0, \ldots, M_2\}$ and as crosses the zeroth order theoretical probabilities $P_{ab}$ 
in Eq.~(\ref{cornerProbs}).  The probabilities marked as circles and crosses have been multiplied by $M_1$ (top edge or bottom edge) or $M_2$ (left-hand edge or 
right-hand edge) for comparison with the functions $f_1^{\rm eff}$ to $f_4^{\rm eff}$.  
Parameters in the left hand plot (a) for the approximate line densities $f_1^{\rm eff}$ to $f_4^{\rm eff}$ 
are given in Eq.~(\ref{numericScaledParametersA}) and for the numerical simulation are as in the caption to Figs.~\ref{fig:StationaryPersp}(a) and (b).  In 
the right hand plot (b) the the mutation and migration rates have been decreased by a factor of 10.  That is, as in Eq.~(\ref{numericScaledParametersC}) for the 
approximate line densities and $M_1 = 100$, $M_2 = 50$, $u_{12} = 0.5 \times 10^{-5}$, $u_{21} = 1 \times 10^{-5}$, $v_{12} = 8 \times 10^{-5}$, $v_{21} = 0.2 \times 10^{-5}$ 
for the numerical simulation.
} 
\label{fig:EdgeEffectivePlots}
\end{center}
\end{figure}

In Fig.~\ref{fig:EdgeEffectivePlots} the effective line densities of Eq.~(\ref{fEffScaledRates}) are compared with the numerically determined stationary 
eigenvector of the transition matrix Eq.~(\ref{transitionMatrix}) along the 4 edges of the square lattice $\{0, \ldots, M_1\}\times\{0, \ldots, M_2\}$.  
Figure~\ref{fig:EdgeEffectivePlots}(a) corresponds to the same set of scaled parameters as Fig.~\ref{fig:TraversePlotsA}, namely Eq.~(\ref{numericScaledParametersA}).   
Figure~\ref{fig:EdgeEffectivePlots}(b) corresponds to scaled mutation and migration rates which are reduced by a factor of 10, namely 
\begin{equation}
\begin{split}
q_{12} = 0.75 \times 10^{-3}, \qquad q_{21} = 1.5 \times 10^{-3}\\
m_{12} = 12 \times 10^{-3}, \qquad m_{21} = 0.3 \times 10^{-3}.  
\end{split}		\label{numericScaledParametersC}
\end{equation}

Recall that Eq.~(\ref{LambdaRestriction}) must be satisfied for the effective line density to be independent of $\Lambda$ to lowest order in $\theta$.  For the parameters 
in Fig.~\ref{fig:EdgeEffectivePlots}(a) we have $\theta \sim 0.1$ and $\Lambda = 1/M_2 =1/50$ (for the top and bottom edges of $\Omega$) or $\Lambda = 1/M_1 = 1/100$ 
(for the left-hand and right-hand edges of $\Omega$).  Thus $\theta |\log \Lambda| \sim 0.5$ and the effective line density is only a coarse approximation to the probability 
that a site will be biallelic throughout the entire population on one island and non-segregating throughout the entire population on the other island.  
In Fig.~\ref{fig:EdgeEffectivePlots}(b) we have $\theta |\log \Lambda| \sim 0.05$, and effective line density is in much closer agreement.  
Note that the asymmetry in each plot is due to a net flux of probability anticlockwise around the boundary of $\Omega$ (see Eq.~(\ref{netFlux})), 
reflecting the asymmetry in migration rates, $m_{12} \ne m_{21}$. 
Also plotted in Fig.~\ref{fig:EdgeEffectivePlots} are the fixation probabilities $P_{ab}$ 
at the corners of $\Omega$ calculated from Eq.~(\ref{cornerProbsScaledRates}), 
which agree poorly with the numerical stationary distribution at the corners in (a) but agree well in (b).  

%
%
\section{Conclusions}
\label{sec:Conclusions}

We have investigated the stationary distribution of the diffusion limit of the 2-island, 2-allele Wright-Fisher model in the limit of small migration and mutation rates.  
By ``small rates'' we mean that the scaled migration rates $m_{ij}$ and mutation rates $q_{ab}$ defined by Eq.~(\ref{mAndQDefn}) 
are assumed to be of the order of a small positive parameter $\theta << 1$.  
An empirical situation relevant to this parameter regime is gene flow between divergent species \citep{souissi2018genomic,stuglik2016genomic}.

Our results for the leading-order-in-$\theta$ stationary distribution near the boundary of its sample space $\Omega$,  
illustrated in Fig.~\ref{fig:UnitSquare}, and for the induced effective line densities and effective point masses at the edges and corners of $\Omega$ are summarised in 
Eq.~(\ref{f1ToF4ScaledRates}) to (\ref{cornerProbsScaledRates}).   

In an infinite sites model the induced effective line densities are site frequency spectra of sites which are bi-allelic in one island and non-segregating in the other.  
Of particular interest is the observation that these line densities include an asymmetric part proportional 
to $m_{12} - m_{21}$, corresponding to a net flux of probability around the perimeter of $\Omega$.  This result is the analogue of similar observation for the 
case of multi-allelic neutral diffusion in a single population, in which the corresponding line densities contain an asymmetric part driven by the non-reversible part 
of the instantaneous mutation rate matrix~\citep{burden2016approximate}.  

The corner probabilities $P_{ab}$ summarised in Eq.~(\ref{cornerProbsScaledRates}) represent the joint probability that a single individual selected at 
random from island 1 is of allele type $A_a$, and a single individual selected at random from island 2 is of allele type $A_b$.  In an infinite sites model, they 
are the relative abundances of allele combinations at sites which are simultaneously non-segregating on both islands.  As expected, for non-zero mutation rates 
allele-type abundances are positively correlated, and for zero mutation rates they are uncorrelated.  Results for the corner probabilities were verified directly from 
the backward generator in Section~\ref{sec:AlternativeFromBackward} and from the coalescent in Section~\ref{sec:AlternativeFromCoalescent}.  The coalescent 
calculation is informative.  It demonstrates that the lowest order approximation to $P_{12}$ and $P_{21}$ corresponds to an ancestry of two individuals, one 
chosen randomly from each island, such that the ancestry includes precisely one migration event occurring before any mutations since the common ancestor, 
followed by any allowed number of mutations since the migration.  

The method described in this paper can in principle be extended to the general case of $g$ islands and $K$ allele types.  In this case the sample space of the stationary distribution 
is a product $\Omega = {\cal S} \times \cdots \times {\cal S}$ of $g$ copies of the $(K - 1)$-dimensional simplex $\cal S$.  There are $K^g$ corner probabilities to determine 
labelled $P_{a_1 \ldots a_g}$, where $a_i = 1, \ldots, K$ labels the allele type fixed on island-$i$.  Each of these probabilities is of $O(1)$.  Corresponding to the states 
in which one island is bi-allelic and the remaining $g - 1$ islands are non-segregating, there are $g \times {k \choose 2} \times K^{g - 1}$ effective line densities, 
contributing a total probability of $O(\theta)$.   The factor $g$ comes from the choice of segregating island;  the factor $K \choose 2$ comes from the choice of two allele types contributing to the bi-allelic site; and the factor $K^{g - 1}$ comes from the allele types fixed on the remaining 
$g - 1$ islands.  The remaining states, i.e.\ those for which more than 
one island is segregating or for which a site is tri-allelic or higher, contribute a total probability of $O(\theta^2)$.  

%
%
\section*{Acknowledgements}

This research was done when Robert Griffiths visited the Mathematical Sciences Institute, Australian National University in November and December 2017. 
He thanks the Institute for their support and hospitality.

\section*{References}
\bibliographystyle{elsarticle-harv}\biboptions{authoryear}





\end{document}